\documentclass[12pt]{article}

\title{Dualism of the $4f$ electrons and high-temperature antiferromagnetism of the heavy-fermion compound YbCoC$_2$}

\usepackage{graphicx}
\usepackage{tabularx}
\usepackage{authblk}
\usepackage{chngpage}
\usepackage{natbib, hyperref}
\usepackage{doi}
\usepackage{amsmath}
\usepackage{chapterbib}

\date{}

\begin{document}

\author[1,2]{D.~A.~Salamatin}
\author[3]{N.~Martin}
\author[1]{V.~A.~Sidorov}
\author[1,4]{N.~M.~Chtchelkatchev}
\author[1,5]{M.~V.~Magnitskaya}
\author[1,5]{A.~E.~Petrova}
\author[1]{I.~P.~Zibrov}
\author[1]{L.~N.~Fomicheva}
\author[6]{Jing Guo}
\author[6]{Cheng Huang}
\author[6]{Liling Sun}
\author[1]{A.~V.~Tsvyashchenko}

\affil[1]{Institute for High Pressure Physics, Russian Academy of Sciences, 14 Kaluzhskoe shosse, 108840 Troitsk, Russia}
\affil[2]{Joint Institute for Nuclear Research, 141980 Dubna, Moscow Region, Russia}
\affil[3]{Universit\'e Paris-Saclay, CNRS, CEA, Laboratoire L\'eon Brillouin, 91191 Gif-sur-Yvette, France}
\affil[4]{Ural Federal University, 19 Mira str., 620002 Ekaterinburg, Russia}
\affil[5]{P.N. Lebedev Physical Institute of the Russian Academy of Sciences, 53 Leninskiy prospekt, 119991 Moscow, Russia}
\affil[6]{Institute of Physics, Chinese Academy of Sciences, 100190 Beijing, P.~R.~China}

\maketitle

\begin{abstract}
We report on the first study of the noncentrosymmetric ternary carbide YbCoC$_2$. Our magnetization, specific heat, resistivity and neutron diffraction measurements consistently show that the system behaves as a heavy-fermion compound, displaying an amplitude-modulated magnetic structure below the N\'eel temperature reaching $T_{\mathrm{N}}$ = 33 K under pressure. Such a large value, being the highest among the Yb-based systems, is explained in the light of our \textit{ab initio} calculations, which show that the $4f$ electronic states of Yb have a dual nature -- i.e.,
due to their strong hybridization with the $3d$ states of Co, $4f$ states expose both localized and itinerant properties.
\end{abstract}


The dual nature of the $5f$ and $4f$-electrons, i.e. the coexistence of localized and itinerant states in a variety of actinide and rare-earth (RE) heavy-fermion systems (e.g., UPt$_3$, UPd$_2$Al$_3$~\cite{Takahashi1996}, UGe$_2$~\cite{Troc2012,Yaouanc2002,Haslbeck2019}, PuCoGa$_5$~\cite{Booth2012}, YbRh$_2$Si$_2$~\cite{Danzenbacher2011}, YbAl$_3$~\cite{Ebihara2000}) is the subject of an intense discussion. A theory for the electronic excitations in uranium compounds, based on the localization of two $5f$ electrons, was developed by Zwicknagl and Fulde~\cite{Zwicknagl2003}.
The standard model for $4f$-electrons in RE metals and their compounds is that an integral number of $4f$ electrons are assumed being
localized at each RE ion site. This applies to both ``normal" REs, where a single $4f$ configuration of given occupation is stable, and to mixed-valent cases, where fluctuations between states with different integral $4f$ occupations take place. The nature of magnetism (itinerant or localized) and the competition between ordered and disordered ground states in these mixed-valent (``abnormal") RE elements (Ce, Eu and Yb), remain a major challenge in condensed matter physics. In this context, the Kondo effect and the Ruderman--Kittel--Kasuya--Yosida (RKKY) interaction arising between itinerant and localized $4f$-electrons play essential roles, as originally pointed out by Doniach~\cite{Doniach1977}.

The noncentrosymmetric carbide YbCoC$_2$, first synthesized more than 30 years ago~\cite{Jeitschko1986}, represents an interesting platform
for studying such physics. Here, we report on \emph{the first study} of its bulk magnetic and transport properties at ambient and
elevated pressure, complemented by neutron diffraction measurements and \textit{ab initio} calculations. We show that YbCoC$_2$ is a heavy-fermion compound with an amplitude modulated incommensurate magnetic order. Its N\'{e}el temperature is the highest among Yb$^{3+}$-based systems,
exceeding the previous record value belonging to $\beta$-YbAlB$_4$~\cite{Tomita2016}. Taken together, our experimental and numerical results indicate that the dual nature of the $4f$ electrons is essential for understanding the magnetic and transport properties of YbCoC$_2$.


The transition from localized to itinerant electron states signalling the onset of large-range electronic correlations,
is also of great interest.
In lanthanides this effect is related to Ce, the first element with a single $4f$ electron level.
The Yb intermetallic compounds are usually considered as being dominated by two valence states, where Yb$^{3+}$ ions can be seen as $f$-hole analogs of Ce.
Owing to such an electronic configuration, one can anticipate a dual character of $4f$ states in YbCoC$_2$, with transitions from
localized states to itinerant ones. Heavy-fermion magnetism in Yb-based systems with localized $4f$ electrons mainly occurs
in compounds where the ordering temperatures is low (typically, $T_{\textrm{N}} < 10~\text{K}$),
for instance in YbNi$_2$~\cite{Rojas2012}, YbNiSn~\cite{Kasaya1992}, YbPtGe~\cite{Katoh2009,Tsujii2015}, YbPdGe~\cite{Tsujii2015, Enoki2012} or YbRh$_2$Si$_2$~\cite{Gegenwart2002}, YbPtBi~\cite{Fisk1991, Ueland2014}.
In fact, many such compounds displaying antiferromagnetic (AFM) order, $T_{\mathrm{N}}$ is even below 1~K (e.g., YbIr$_2$~\cite{Willis1985}, YbPd~\cite{Pott1985}, Yb$X$ ($X$ = N, P, As)~\cite{Ott1985} and Yb$X$Cu$_2$ ($X$ = Au, Pd)~\cite{Rossel1987}).
[Few exceptions are YbBe$_{13}$ ($T_{\mathrm{N}}$ = 1.3~K)~\cite{Walter1985}, Yb$_3$Pd$_4$ ($T_{\mathrm{N}}$ = 3~K)~\cite{Pollit1985} and YbB$_2$ ($T_{\mathrm{N}}$ = 5.7~K)~\cite{Avila2003}.]
At atmospheric pressure, the highest magnetic transition temperature is $T_{\mathrm{N}} = 20$~K observed recently in $\alpha$-YbAl$_{1-x}$Mn$_{x}$B$_{4}$ for $x$ = 0.27~\cite{Suzuki2018}.
The pressure-induced magnetic ordering was found in the heavy-fermion superconductor $\beta$-YbAlB$_4$~\cite{Tomita2015a}, where the magnetic transition temperature increases up to 32~K, obtained under the external pressure $P$ of 8~GPa~\cite{Tomita2016}.

\begin{figure}[ht]
\includegraphics[width=\textwidth]{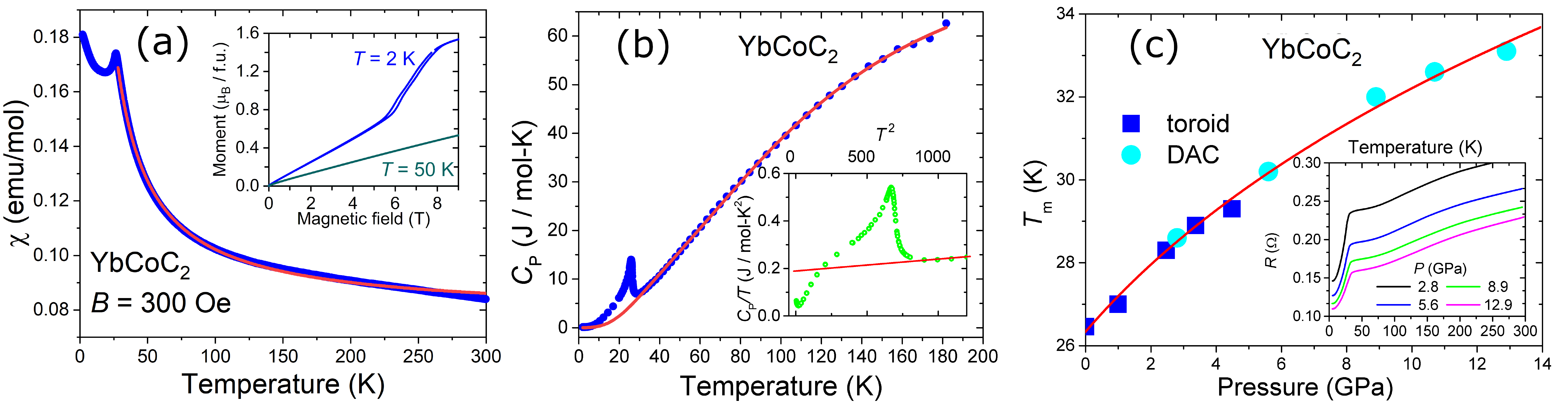}
\caption{\label{ExperMagnet}(a) Temperature-dependence of the magnetic susceptibility $\chi$ of YbCoC$_2$, measured in the
external magnetic field $B$=300~Oe (blue points). The red line is the fit of a modified Curie-Weiss law to the $T$=28--300~K data 
(see the text for details). Inset: isothermal magnetization of YbCoC$_2$ measured in the $B$ = 0--9~T field range at $T$ = 2 and 50~K. (b) Temperature-dependence of the heat capacity $C_{\rm P}$ of YbCoC$_2$ (blue points). The red line is the fit of two Debye laws to the high-temperature data, extrapolated down to $T \rightarrow 0$ (see the text). Inset: the low-temperature part of $C_{\rm P}/T$ vs. $T^2$.
The red line is the linear approximation of $C_{\rm P}/T$ vs. $T^2$ extended down to 0~K. (c) Pressure-dependence of the magnetic transition temperature $T_{\rm m}$ measured in the toroid and diamond-anvil (DAC) high-pressure cells (blue squares and light blue circles, respectively).
The red line is the approximation with the function $f (P - P_{c0})^{\alpha}$, where $f$ = 18.6~K/GPa$^{1/5}$, $P_{c0}$ = -5.7~GPa and $\alpha$ = 1/5. Inset: temperature-dependence of the isobaric electrical resistance, measured at $P$ = 2.8--12.9~GPa.}
\end{figure}

Previously it has been established that the magnetic structure of the $R$CoC$_2$ compounds (where $R$ is a heavy RE element), isostructural to YbCoC$_2$, is ferromagnetic (FM). The magnetic moments usually point along the $a$ axis~\cite{Schafer1997}, except in DyCoC$_2$ where they lie in the $ac$ plane~\cite{Amanai1995}. The magnetic susceptibility measurements in YCoC$_2$ and the refinements of neutron powder diffraction patterns of other $R$CoC$_2$ compounds indicate that the Co and C ions are non-magnetic~\cite{Schafer1990, Amanai1995}. Notably, the Curie temperature for $R$CoC$_2$ compounds deviates from the de Gennes law~\cite{DeGennes1962,Schafer1997}, indicating the importance of the interactions
beyond RKKY in the stabilization of the magnetic order in this series and calls for further studies of the nature of magnetism in these systems.


Polycrystalline samples of YbCoC$_2$ were synthesized by melting Yb, Co and C (see Ref.~\cite{Tsvyashchenko1984}) at 8~GPa and 1500-1700~K using Toroid high-pressure cell. The Rietveld analysis of x-ray and neutron diffraction data at $T$ = 300~K shows that the compound crystallizes in an orthorhombic structure of the CeNiC$_2$-type (space group $Amm2$, no. 38), similar to other RE carbides $R$CoC$_2$. High-pressure X-ray diffraction measurements performed at the 15U beamline of the Shanghai Synchrotron Radiation Facility (China) show that the orthorhombic crystal structure is preserved up to 37 GPa~\cite{suppl}. A fit of a Murnaghan equation of state volume vs. pressure data yields a bulk modulus $B_0 = 176(23)$~GPa, with its first pressure derivative $B^{\prime}_0 = 9(1)$.
The bulk magnetic properties of YbCoC$_2$ were studied by vibrating sample magnetometry (VSM) using a physical property measurement system (PPMS). The electrical resistivity measurements were performed on bulk polycrystalline samples using a lock-in detection technique~\cite{Sidorov_HPcell}. The high-pressure resistance
measurements below 5 GPa were performed in a miniature clamped Toroid-type device with glycerine--water (3:2) liquid as the pressure transmitting medium. For higher pressures a diamond-anvil cell (DAC) with NaCl as the pressure transmitting medium was used.


The magnetic susceptibility $\chi(T)$ measured in the magnetic field $B$ = 300~Oe exhibits a sharp peak, indicative of an AFM-like transition at $T_{\mathrm{N}}$ = 25.6(2)~K (Fig.~\ref{ExperMagnet}a). In the $T = 28$--$300$~K-range, $\chi(T)$ can be well-described by a modified Curie--Weiss law $\chi = \chi^*_0 + C^* / \left(T-\theta\right)$, where $\chi^*_0$ is a temperature-independent term, $C^*$ - the Curie constant and $\theta$ - the Weiss constant. The effective magnetic moment deduced from this approximation is $\mu_{\mathrm{eff}} = 4.31(1)~\mu_{\mathrm{B}}$ per formula unit (f.u.), a value close to that of the free Yb$^{3+}$ ion ($4.54~\mu_{\mathrm{B}}$). The small positive Weiss constant $\theta$ = 2.4(1)~K reveals the presence of weak ferromagnetic correlations and a small value of the ratio $f_s = \theta/T_{\mathrm{N}} \approx  0.09 << 1$ indicates a weak level of spin frustration in the system.

The inset of Fig.~\ref{ExperMagnet}a shows field-dependences of the bulk magnetization $M$ measured in the $B$ = 0--9~T-range at $T$ = 2 and 50~K. At $T$ =50~K, $M(H)$ has a smooth linear character, typical of a paramagnet. In contrast, the curve measured at $T$ = 2~K $< T_{\mathrm{N}}$ first increases linearly up to $\approx$ 6 T, until a field-induced ferromagnetic transition with a hysteresis of about 0.2~T takes place in the $\approx$ 6--8~T-range. $M$ does not fully saturate at the highest field of 9 T and reaches a value of $\approx$ 1.6~$\mu_{\mathrm{B}}$/f.u., which stands well-below the theoretical saturation moment for the full Yb$^{3+}$ multiplet ($\mu_{\mathrm{s}} = 4.0~\mu_{\mathrm{B}}$).

\begin{figure}[ht]
\includegraphics[width=\textwidth]{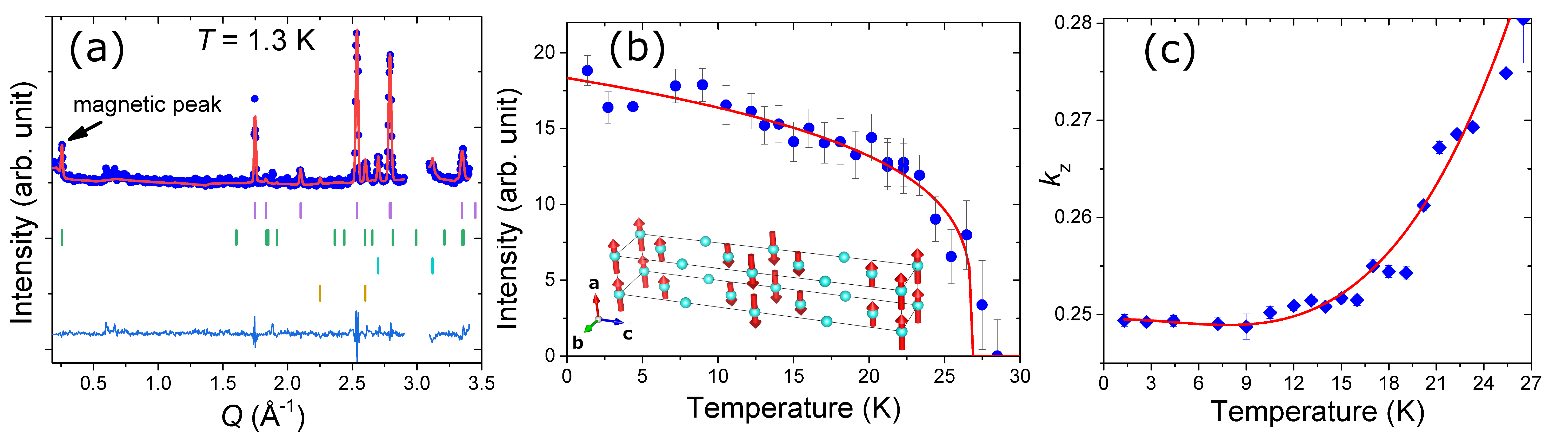}
\caption{\label{ExperNeutron}(a) Refined neutron powder-diffraction pattern of YbCoC$_2$ measured at $T$ = 1.3~K and ambient pressure: the experimental points (blue dots), the calculated profiles (red line) and their difference (light blue line).
The bars in the lower part of the graph represent the calculated Bragg reflections that correspond to the nuclear $Amm2$ structure (first row, purple), the magnetic phase (second row, green), the aluminum container (third row, cyan), and the non-magnetic impurity phase YbO (fourth row, brown). (b) Temperature-dependence of the square-root of the integrated intensity of the magnetic peak (blue circles).
The red line is the power law fit (see the text). Inset: Schematic representation of the magnetic unit cell of YbCoC$_2$ in the commensurate phase (only Yb magnetic atoms are shown). (c) Temperature-dependence of the $k_z$ component of the magnetic wave vector (blue diamonds).
The red line is drawn to guide the eye.}
\end{figure}
%


Fig.~\ref{ExperMagnet}b displays the temperature-dependence of the heat capacity $C_{\mathrm{P}}$, obtained in the 2-180~K range. Around $T_{\mathrm{N}}$, the $C_{\mathrm{P}}(T)$ curve exhibits a lambda anomaly, with a maximum at $\approx$~26~K, which is consistent with the bulk AFM-like transition deduced from the $\chi(T)$ data. The heat capacity jump at $T_{\mathrm{N}}$ is equal to $\Delta C_{\mathrm{P}}(T_{\mathrm N})$ = 14.64~J/mol-K. This value is close to the one calculated for an amplitude modulated magnetic structure for Yb$^{3+}$, namely 13.43~J/mol-K~\cite{Blanco1991}. The heat capacity at $T >$ 30~K was approximated by $C_{\mathrm{fit}}(T) = m_1D(T,\Theta_1) +  m_2D(T,\Theta_2)$, where $D$ is the Debye function, with the characteristic temperatures $\Theta_1 = 516(11)$~K and $\Theta_2 = 176(7)$~K. The best fit was obtained with the coefficients $m_1$ = 2.49(4) and $m_2$ = 0.79(5).
The magnetic entropy of the transition computed according to $S_{\mathrm{m}}(T) = \int_{\mathrm{1.9~K}}^{T} C_{\mathrm{m}}(T) / T \, dT$~, where $C_{\mathrm{m}}(T) = C_{\mathrm{P}}(T) - C_{\mathrm{fit}}(T)$ is the magnetic part of heat capacity, amounts to 3.91~J/mol-K at $T$ = 30~K. This value is about 70~\% of that expected for a magnetic doublet ground state and is most likely due to Kondo screening \cite{Blanco1994}. The Sommerfeld coefficient obtained from the $C_{\mathrm{P}}$ data just above $T_{\mathrm{N}}$, is $\gamma = 190(1)$~mJ/mol-K$^2$ (see inset of Fig.~\ref{ExperMagnet}b). This value suggests a considerable enhancement of the effective electron mass in YbCoC$_2$
comparable with other heavy-fermion compounds~\cite{Katano2017, Motoya1997}.
The low temperature upturn in $C_{\mathrm{P}}/T$ (see the inset of Fig~\ref{ExperMagnet}b) indicates that there is a possibility of the second magnetic transition in YbCoC$_2$ similar to the incommensurate-commensurate magnetic transition in CeNiC$_2$ \cite{Motoya1997}.


As shown in Fig.~\ref{ExperMagnet}c, the electrical resistance increases monotonically with temperature revealing the metallic character of YbCoC$_2$.
Because of the coherent scattering of charge carriers in the magnetically ordered state,
the resistance drops down below the AFM-like transition temperature (see inset of Fig.~\ref{ExperMagnet}c). Interestingly, the transition temperature $T_{\rm m}$, determined from the onset of the resistance, increases with external pressure up to 33.2(3) K at $P$ = 13~GPa.

\begin{figure}[!ht]
  \centering
  \includegraphics[width=0.8\columnwidth]{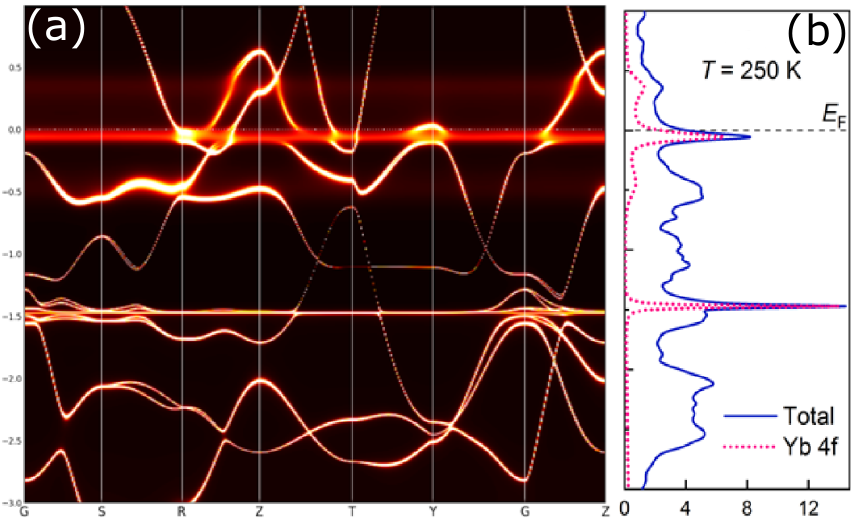} 	   \includegraphics[width=0.5\columnwidth]{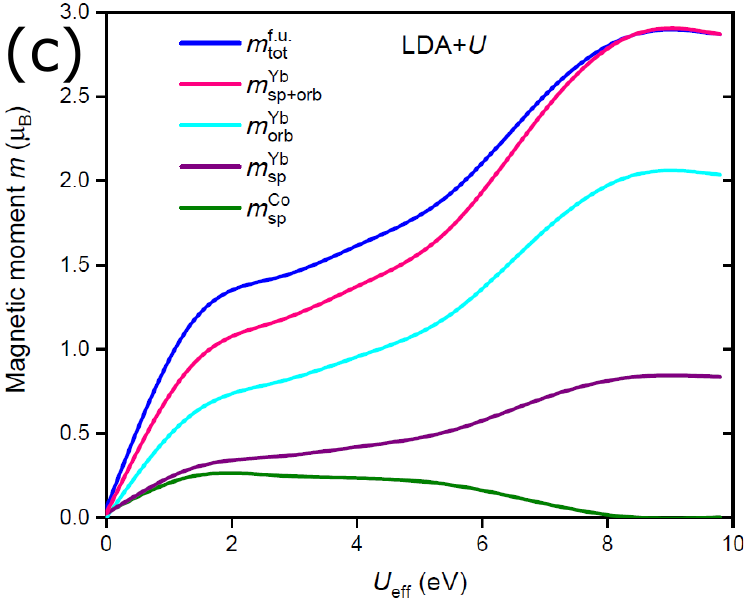}
  \caption{The DMFT band structure (a) and DOS (b) at $T$ = 250 K. The DOS is in arb. units and the Fermi energy is set to zero. (c) The LDA+$U$ calculations: the magnetic moment per formula unit and its partial contributions as functions of $U_{\mathrm{eff}}$. The green line is the spin moment of Co, the violet line is the spin moment of Yb, the cyan line is the orbital moment of Yb, the red line is the total moment of Yb and the blue line is the total magnetic moment.} \label{fig:TheorFigs}
\end{figure}

%


In order to elucidate the magnetic structure of YbCoC$_2$, we have used neutron powder diffraction (NPD) on the thermal instrument G4-1 (Laboratoire L\'eon Brillouin). The low temperature patterns display an additional peak at low angles (see Fig.~\ref{ExperNeutron}a). The square-root of the integrated intensity of this peak depends on temperature (see Fig.~\ref{ExperNeutron}b) and may be approximated with the function $\sigma_{0}(1 - T/T_{\mathrm{C}})^{\beta}$, with the critical exponent $\beta = 0.24(4)$ and the critical temperature $T_{C} = 26.9(7)$~K. This peak can be indexed with the wave vector $(0, 0, k_z)$, where $k_z$ is temperature-dependent (see Fig.~\ref{ExperNeutron}c).
The best refinement of the magnetic part of neutron powder diffraction pattern is achieved by assuming an amplitude modulated (sine-wave) structure of Yb$^{3+}$ magnetic moments, pointing along the $a$ axis with the amplitude 1.32(7)~$\mu_{\mathrm{B}}$ at $T=1.3$~K.
Such a value of magnetic moment is typical for Yb in magnetic compounds~\cite{Tsujii2015, Chattopadhyay2018, Yaouanc2016, Ueland2014} and may be attributed to the crystal electric field effects and a partial screening due to the Kondo mechanism. The structure locks into a commensurate state with $k_z = 1/4$ at $T_{\mathrm{lock-in}} \approx 8$~K.


The \textit{ab initio} simulations of YbCoC$_2$ were performed using the Wien2k package~\cite{wien2k} within the density functional theory (DFT) and the local density approximation (LDA), with the spin-orbit coupling (SOC) taken into account. The $4f$ electrons of the Yb ion were explicitly treated as valent and not completely localized, and as such were allowed to and hybridize with all other states. The calculations started from the experimental lattice parameters measured in this work, with subsequent relaxation of internal atomic coordinates. In these initial DFT-LDA calculations ($T$ = 0~K), a nonmagnetic (paramagnetic) ground state of YbCoC$_2$ was determined, whereas no magnetic solution turned out
to be stable.

From the band structure and density of states (DOS) of paramagnetic YbCoC$_2$ it follows~\cite{suppl} that $4f$ states of Yb are present
at the Fermi energy ($E_{\mathrm{F}}$), and the main contribution to the total DOS at $E_{\mathrm{F}}$
stems from the strongly hybridized ytterbium $f$- and cobalt $d$-states.
The energy bands in the vicinity of $E_{\mathrm{F}}$ are flat, which indicates a strong enhancement of the electron mass
to a value typical for heavy-fermion systems.

To treat strong electron correlations in the Yb $4f$ shell adequately, we further employed the combination of DFT-LDA~\cite{wien2k} and dynamical mean field theory (DMFT)~\cite{Kotliar2006} as implemented in the eDMFT package~\cite{Haule2010}.
Since the DMFT method is not applicable to the low-temperature region, for simulations of the magnetic state of YbCoC$_2$ observed
experimentally below 30~K we have used the LDA+$U$ approach~\cite{Anisimov1993,Liecht1995} in the Wien2k implementation,
which can perform zero-temperature calculations with accounting for correlation effects of $4f$ electrons.

The band structure and DOS computed at $T$ = 250~K using DMFT are shown in FIG.~\ref{fig:TheorFigs}(a, b). The evaluated number of $4f$ electrons is about 13.05, i.e. the valence of Yb is actually equal to three. There is a certain similarity between the DFT and DMFT results in the number of $4f$ electrons and in the position of the bands and corresponding DOS peaks~\cite{suppl}. Within both approaches, the Fermi energy falls into a steep slope of the $4f$ DOS whose maximum is very close to $E_{\mathrm{F}}$.
Apparently, the proximity of $4f$ peak to $E_{\mathrm{F}}$ and the resulting relatively high $N(E_{\mathrm{F}})$,
as well as a significant $f$--$d$ hybridization favour the high magnetic transition temperature for YbCoC$_2$.

In our LDA+$U$ simulations of the magnetic state only the $4f$ electron correlations were taken into account.
The positions of the bands are found to shift downwards with increasing $U_{\mathrm{eff}} = U - J$ (where $U$ and $J$ are the standard on-site Coulomb interaction constants). We obtained a stable FM ordering with the periodicity of the crystal lattice.
The total magnetic moment and its main partial atomic contributions as functions of $U_{\mathrm{eff}}$ are depicted in Fig.~\ref{fig:TheorFigs}(c).
The largest partial contribution (including both the spin and orbital terms) is $\mu_{\mathrm{Yb}}$ directed along the $a$ axis in
accordance with the experimental data and drawn in Fig.~\ref{fig:TheorFigs}(c) as the red line.
This is the magnetic moment of the Yb site which increases with $U_{\mathrm{eff}}$.
The magnetic moment of Co is smaller tending to zero with increasing $U_{\mathrm{eff}}$, whereas the magnetic moments at carbon sites and in the interstitial region are negligible at any $U_{\mathrm{eff}}$.

The interval of $U_{\mathrm{eff}}$ between 1.3 and 3.5 eV has been considered as optimal, corresponding to the experimental situation.
The change in $U_{\mathrm{eff}}$ in this interval does not lead to a strong variation of the magnetic moments and band positions.
The Yb moment varies from 0.9 to 1.3 $\mu_{\mathrm{B}}$, which correlates well with the experimental observations. The model of ferromagnetic alignment used in the calculations turned out to be quite reasonable, because the period of experimental modulated magnetic structure is about four times that of the crystal lattice. Moreover the Yb magnetic moments are co-directed to each other within the crystallographic unit cell (see the inset in Fig.~\ref{ExperNeutron}b). Thus, our LDA+$U$ calculations demonstrate that YbCoC$_2$ possesses a stable magnetic ordering with the magnetization predominantly located at the Yb sites.


A wealth of experimental and numerical results presented in this study, unambiguously establishes that the noncentrosymmetric YbCoC$_2$ is an unusual heavy-fermion system, displaying incommensurate antiferromagnetic ordering
with the transition temperature reaching 33~K under pressure.
The nature of its magnetic structure is surprising in itself, since the isostructural compounds $R$CoC$_2$ (where $R$ = Gd--Tm)
display FM ordering. This fact is well captured by our {\it ab initio} calculations, which also reproduce the direction of the Yb magnetic moments and, quantitatively, their amplitude. As a key result, we reveal a strong hybridization between the $4f$ states of Yb and the $3d$ orbitals of Co. We believe that this feature is instrumental in explaining the robustness of the magnetic order and its enhancement under applied external pressure.
Another appealing finding concerns the evolution of the magnetic structure with pressure. Indeed, a fine-tuning of the tight competition between RKKY and Kondo interactions is a well-proven route towards the observation of quantum criticality, exotic forms of superconductivity and novel strange metallic states.

\section*{Acknowledgement}
This work was supported by Russian Science Foundation: D.A.S., V.A.S., A.E.P., and A.V.T. acknowledge the support of their experimental work (grant by the RSF No. 17-12-01050); N.M.C. and M.V.M. are grateful for support of their theoretical calculations (Grant RSF 18-12-00438). The results of calculations were obtained using computational resources of MCC NRC `Kurchatov Institute` (http://computing.nrcki.ru/) and supercomputers at Joint Supercomputer Center of RAS (JSCC RAS). We also thank the Uran supercomputer of IMM UB RAS for access.

\newpage
\appendix
\begin{center}{\Large\bf Supplementary Information}\end{center}
\begin{abstract}
We present supplementary information on the details of experimental techniques and \textit{ab initio} calculations discussed in the main text.
\end{abstract}

\subsection*{Experimental techniques\label{Sec1}}
The purity of constituent elements of YbCoC$_2$ were 99.7\% for Yb, 99.9\% for Co and 99.99\% for C. 
For high-pressure x-ray diffraction measurements at the 15U beamline of the Shanghai Synchrotron Radiation Facility a monochromatic x-ray beam with a wavelength of 0.6199~\AA~was employed and silicon oil was used as a pressure-transmitting medium. The pressure for all measurements in the DACs was determined by the ruby fluorescence method~\cite{Mao1986}.

X-ray powder diffraction measurements at room temperature and ambient pressure were performed using the diffractometer Guinier camera G670, Huber (Cu-K$\alpha_1$). This XRD pattern was analyzed with GSAS software~\cite{GSAS}. NPD patterns were analyzed with the Rietveld and Le Bail methods using FULLPROF software~\cite{Fullprof}. The measured lattice constants are in agreement with previous determinations~\cite{Jeitschko1986}.

\subsection*{Structure\label{Sec2}}
\begin{figure}[!ht]
  \centering
\includegraphics[width=0.8\columnwidth]{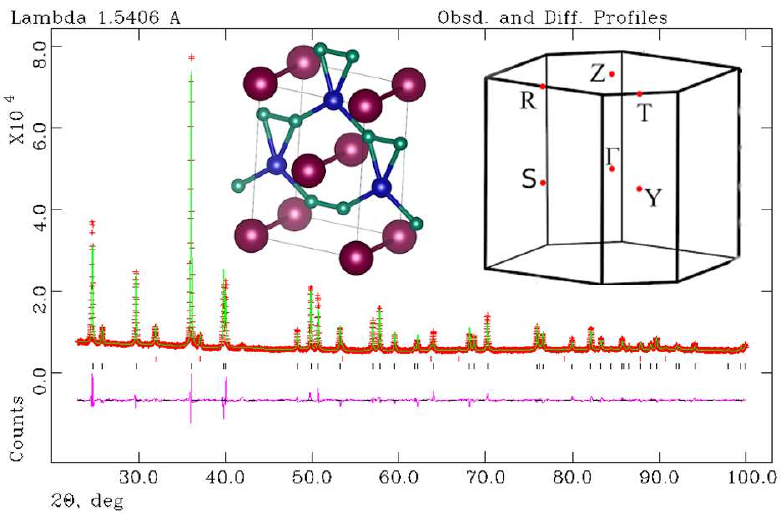}
\renewcommand{\figurename}{Fig. S}
\caption{\label{XRD_patt} XRD pattern for YbCoC$_2$ at room temperature and $P = 0$~GPa (the obtained lattice parameters are listed in Table~\ref{table1}). The experimental points (red marks), the calculated profiles (green line) and their difference (purple line) are shown. The bars in the lower part of the graph represent the calculated Bragg reflections that correspond to the YbO impurity phase (upper row, red color) and YbCoC$_2$ (lower row, blue color). Inset: The unit cell (generated using the VESTA software~\cite{Momma2011}), with the red, blue and green balls corresponding to Yb, Co and C ions (left) and the first Brillouin zone (right) of the $Amm2$ structure of YbCoC$_2$.}
\end{figure}

The unit cell of this structure is shown in together with its first Brillouin zone (right). Planes of Yb ions perpendicular to the $a$ direction alternate with the planes of Co and C ions shifted along the $a$ direction by $a/2$ (see FIG.~S\ref{XRD_patt}). The C ions exist as dimers with the C-C nearest distance of 1.55 \AA, double bonds between them shown with green sticks. Red sticks indicate the nearest distance between two Yb ions equal to the a lattice period (3.44 \AA), which hardly assumes essential direct exchange between localized $4f$ electrons of Yb.
The impurity of high-pressure phase of non-magnetic ytterbium oxied (YbO) with a fraction of less than 5\%~wt. was also found in the samples.

\begin{figure}[!ht]
  \centering
\includegraphics[width=0.8\columnwidth]{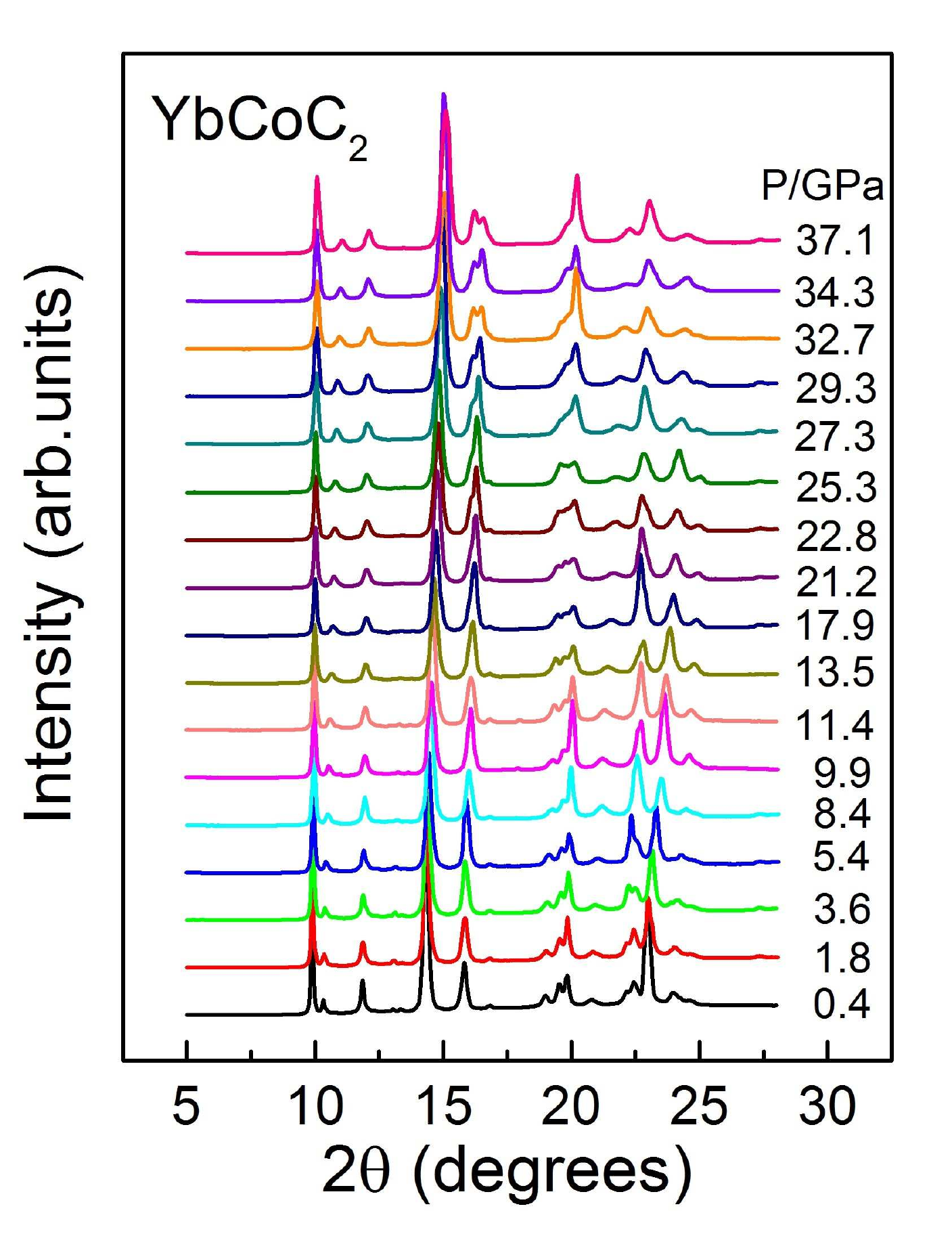}
\renewcommand{\figurename}{Fig. S}
\caption{\label{XRD_HP} X-ray diffraction patterns of YbCoC$_2$ at different pressures.}
\end{figure}

\begin{figure}[!ht]
  \centering
\includegraphics[width=0.8\columnwidth]{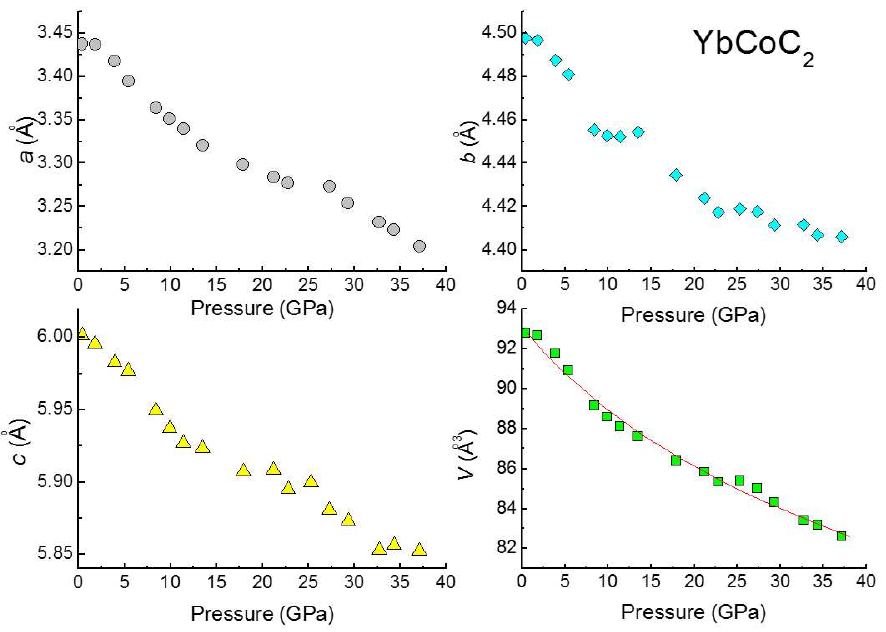}
\renewcommand{\figurename}{Fig. S}
\caption{\label{latt_const_HP} Pressure dependences of the lattice parameters and the unit cell volume of YbCoC$_2$. Red line is the Murnaghan fit of experimental $V(P)$ dependence.}
\end{figure}

X-ray diffraction patterns of YbCoC$_2$ obtained at different pressures are presented in FIG.~S\ref{XRD_HP}. The dependencies of the lattice constants and unit cell volume of YbCoC$_2$ under high pressure up to 37 GPa are presented in FIG.~S\ref{latt_const_HP}.

The results of Rietveld refinements of XRD and NPD patterns at room temperature and atmospeheric pressure are presented in Table~\ref{table1}.

\begin{table}
 \begin{adjustwidth}{-1.5in}{}  
        \begin{center}
\caption{\label{table1} The crystallographic parameters of YbCoC$_2$ (space group $Amm2$) obtained at $T = 300$~K and $P = 0$~GPa}
\begin{tabular}{|c|c|c|c|c|c|c|c|c|}
	\hline
	\multicolumn{9}{|c|}{X-ray diffraction Cu-K$\alpha_1$} \\
	\hline
	$a$ (\AA) & $b$ (\AA) & $c$ (\AA) & $z$(Co) & $y$(C) & $z$(C) & $U_{\mathrm{iso}}$(Yb) (\AA$^{2}$) & $U_{\mathrm{iso}}$(Co) (\AA$^{2}$) & $U_{\mathrm{iso}}$(C) (\AA$^{2}$) \\
	\hline
	3.4391(1) & 4.4948(1) & 5.9987(1) & 0.6146(9) & 0.172(2) & 0.331(1) & 0.0176(2) & 0.0175(8) & 0.0306(34) \\
	\hline
	\multicolumn{9}{|c|}{Neutron diffraction $\lambda = 2.42$~\AA} \\
	\hline
	3.4411(9) & 4.4881(13) & 5.9925(15) & 0.643(24) & 0.154(4) & 0.311(6) & 0.0 & 0.0 & 0.0 \\
	\hline
\end{tabular}
\end{center}
    \end{adjustwidth}
\end{table}

\subsection*{Resistivity\label{Sec3}}
\begin{figure}[!ht]
  \centering
\includegraphics[width=0.8\columnwidth]{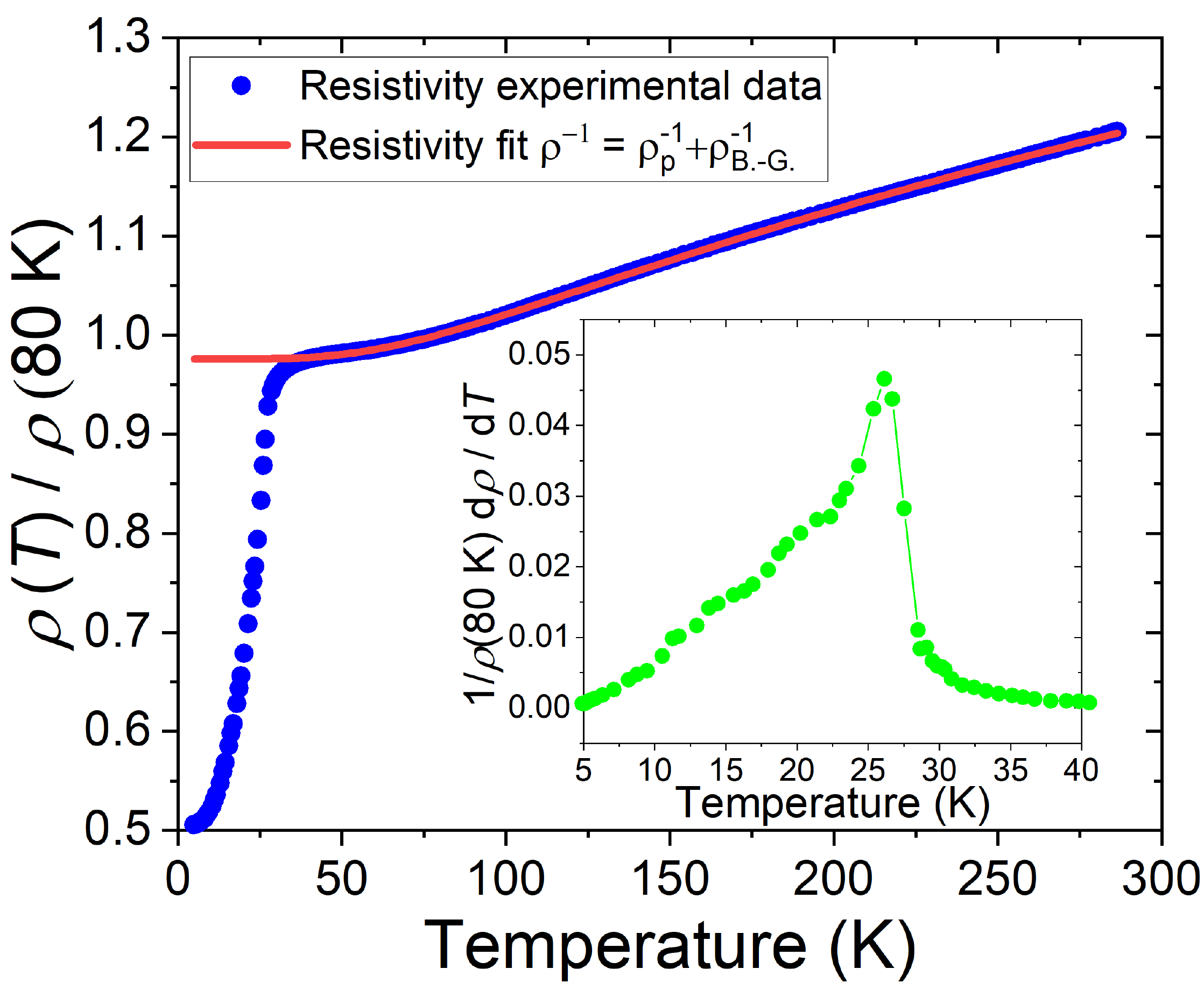}
\renewcommand{\figurename}{Fig. S}
\caption{\label{resistivity_NP} Measured resistivity relative to the value of resistivity at $T = 80$~K for YbCoC$_2$ at $P = 0$~GPa. The absolute value of resistivity at $T = 298$~K is 680 $\mu\Omega$~cm.
Blue points: experimental data, red line: fit with parallel resistor and Bloch-Gruneisen model (see text for details).
Inset: the temperature derivative of resistivity at low temperatures.}
\end{figure}

The relative resistivity above $T = 40$~K could be well fitted with the model $\rho^{-1}_{\mathrm{fit}} (T) = \rho^{-1}_{\mathrm{p}} + \rho^{-1}_{\mathrm{B-G}} (T)$ (see FIG.~S\ref{resistivity_NP}), where $\rho_{\mathrm{p}}$ - temperature-independent parallel resistivity and $\rho_{\mathrm{B-G}} (T) = \rho_0 + A_{\mathrm{el-ph}} T(\frac{T}{\Theta_\mathrm{D}})^4 \int_{0}^{\Theta_{\mathrm{D}}/T} \frac{x^5}{(e^x - 1)(1 - e^{-x})} dx$ where
$\rho_{0}$ - 	residual resistivity and the second term is contribution to the resistivity from electron-phonon scattering approximated by Bloch-Gruneisen law with $n = 5$
($A_\mathrm{el-ph}$ - electron-phonon coupling constant and $\Theta_{\mathrm{D}}$ - Debye temperature).
The best fit was obtained with the following parameters $\rho_{\mathrm{p}} = 2.46(6)$, $\rho_0 = 1.62(2)$, $A_{\mathrm{el-ph}} = 0.015(4)$~K$^{-1}$ and $\Theta_{\mathrm{D}} = 412(3)$~K.
It is interesting to note that Debye temperature determined from the resistivity is close to the average value of Debye temperature obtained from the heat capacity measurements
(from heat capacity $<\Theta_{\mathrm{D}}> = \frac{m_1 \Theta_1 + m_2 \Theta_2}{m_1 + m_2} = 434(5)$~K).
The carefull study of the difference between measured and fitted resistivities did not reveal any anomalies above the antiferromagnetic transition temperature.
The temperature derivative of the resistivity at $P = 0$~GPa near antiferromagnetic tranisition temperature shows a peak with a maximum at $T = 26.1(7)$~K (see inset to FIG.~S\ref{resistivity_NP}).

\subsection*{\textit{Ab initio} calculations\label{Sec4}}
We employed the density-functional-theory (DFT), DFT+$U$, and DFT+DMFT methods to theoretically explore electronic and magnetic properties of YbCoC$_2$, with local density approximation (LDA) for exchange-correlation potential. The spin-orbit coupling (SOC) was taken into account.

Our \textit{ab initio} DFT-LDA simulations were performed using the relativistic APW+lo method as implemented in Wien2k package~\cite{wien2k}. For core electrons, SOC effects were included self-consistently by solving the radial Dirac equation. For valence states, the SOC was evaluated by the second-variational step~\cite{Macdonald1980} using scalar-relativistic eigenvectors. Since the SOC term is large only near the core, the corresponding contributions to the Hamiltonian were only evaluated inside the MT spheres surrounding the atoms. The MT-radii of Yb, Co, and C were set to 2.50, 1.98, and 1.25 $r_{\mathrm{B}}$, respectively. The convergence factor $R_{\mathrm{min}}K_{\mathrm{max}}$ was set to 8.0, where $R_{\mathrm{min}}$ is the minimal MT-radius and $K_{\mathrm{max}}$ is the plane-wave cut-off parameter, $K_{\mathrm{max}}^2 \approx$ 560~eV. This, together with the reciprocal space resolution of $2\pi \times 0.03~\AA^{-1}$ employed to sample the Brillouin zone (BZ) ensured the total energy convergence of about $10^{-7}$~eV. The calculations were made at experimental lattice periods found in this work. Starting from the experimental atomic positions, we did the geometry relaxation to allow the internal atomic coordinates to change, until the residual atomic forces were converged down to 3 meV/$\AA$. The 4$f$ electrons of Yb were explicitly treated as valent.

\begin{figure}[!ht]
  \centering
  \includegraphics[width=0.6\columnwidth]{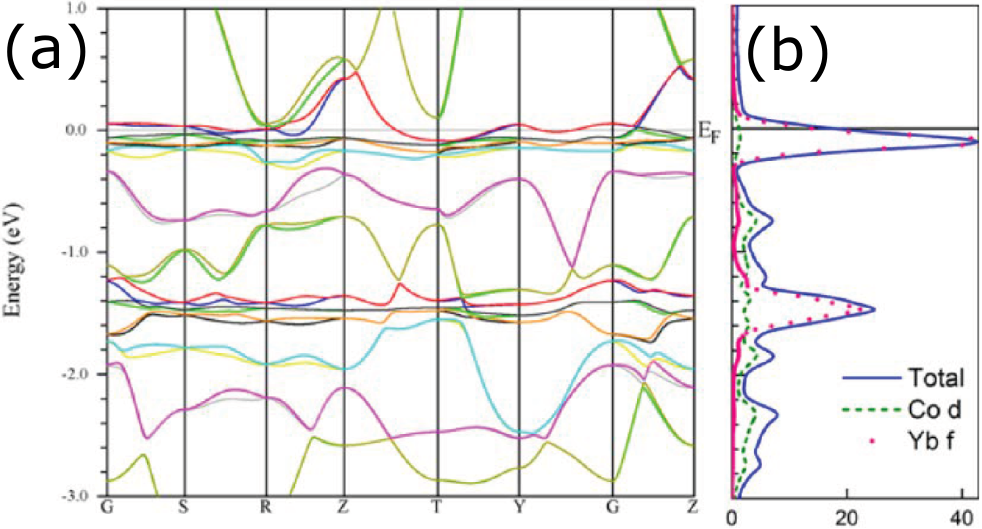} \includegraphics[width=0.4\columnwidth]{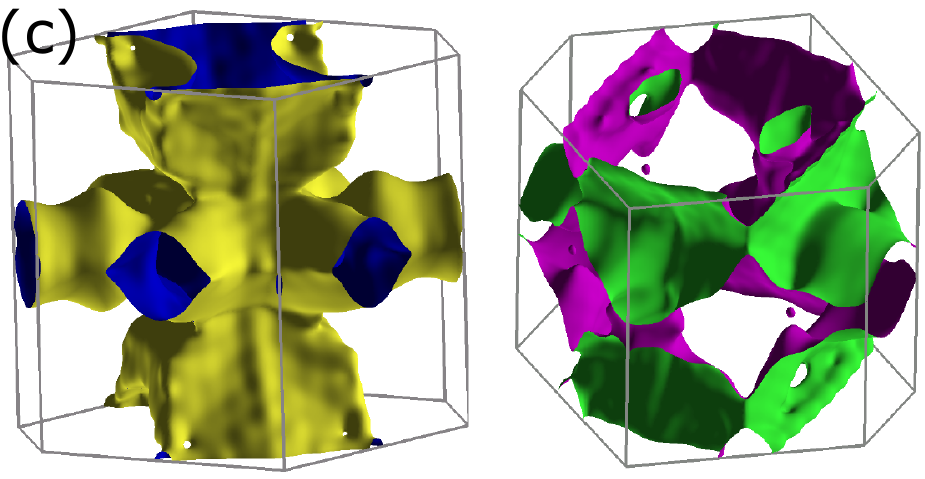}
   \renewcommand{\figurename}{Fig. S}
   \caption{(a, b) The DFT band structure and density of states (DOS), correspondingly; the DOS is mainly contributed by hybridized $4f$-Yb and $3d$-Co states, other contributions are much smaller and not shown. (c, d) The Fermi surface sheets corresponding to the 49th and 50th bands, with standard notation for high-symmetry points of Brillouin zone. The FS was generated using the XCRYSDEN software~\cite{Kokalj2003}.}\label{fig:DFT}
\end{figure}

FIG.~S\ref{fig:DFT}(a,b) presents the band structure and density of states (DOS) calculated within DFT-LDA. The DOS near Fermi energy ($E_{\mathrm{F}}$) is mainly contributed by the hybridized 4$f$-Yb and 3$d$-Co states. The corresponding Fermi surface (FS) topology in the first BZ is shown in FIG.~S\ref{fig:DFT}(c,d), with standard notation for the high-symmetry points. Two FS sheets related to the two spin channels of SOC are formed by the 49th and 50th bands (blue and red lines in the band structure, respectively) that cross $E_{\mathrm{F}}$. The 3rd sheet formed by the 48th band (black line) and consisting of tiny hole-like pockets in the $\Gamma$Z direction is not shown.

In nonmagnetic YbCoC$_2$, a calculated value of the DOS at the Fermi level $N(E_{\mathrm{F}})$ is about of 9~st./eV = 21~mJ/(mol~K$^2$) per formula unit. As in case of other Yb-based metallic compounds, our DFT-LDA simulations (made with experimental lattice periods) demonstrate very high sensitivity of $N(E_{\mathrm{F}})$ to the choice of calculation parameters and numerical accuracy, because $E_{\mathrm{F}}$ falls at a very steep slope of the high $4f$ peak. So our estimate of $N(E_{\mathrm{F}})$ is only indicative.

We used the LDA+DMFT approach~\cite{Georges1996,Kotliar2006} as implemented in the eDMFT package~\cite{Haule2010}. The continuous-time quantum Monte Carlo (QMC-DMFT) impurity solver~\cite{Haule2007} was employed. The magnetic state at low temperatures was studied with the Wien2k package~\cite{wien2k} using the LDA+$U$ method~\cite{Anisimov1993}. We used the rotationally invariant version~\cite{Liecht1995} of this method, with the parameter $U_{\mathrm{eff}} = U - J$.

\bibliographystyle{plainnat}
\bibliography{dualism8}

\end{document}